# Nanoscale Investigation of Microcracks and Grain Boundary Wetting in Press Hardened Galvanized 20MnB8 Steel

M. Arndt[a], P. Kürnsteiner[b,e], T. Truglas[b], J. Duchoslav[c, d], K. Hingerl[c], D. Stifter[c], C. Commenda[a], J. Haslmayr[a], S. Kolnberger[a], J. Faderl[a], H. Groiss[b*]

[a]voestalpine Stahl GmbH, voestalpine-Straße 3, 4020 Linz, Austria

[b]Christian Doppler Laboratory for Nanoscale Phase Transformations, Center for Surface and Nanoanalytics (ZONA), Johannes Kepler University Linz, Altenberger Straße 69, 4040 Linz, Austria

[c]Center for Surface and Nanoanalytics (ZONA), Johannes Kepler University Linz, Altenberger Straße 69, 4040 Linz, Austria

[d]Centre for Electrochemistry and Surface Technology GmbH (CEST), Viktor Kaplan Straße 2, 2700 Wiener Neustadt, Österreich

[e]Department Microstructure Physics and Alloy Design, Max-Planck-Institut für Eisenforschung GmbH, Max-Planck-Straße 1, 40237 Düsseldorf, Germany

*: corresponding author: heiko.groiss@jku.at

## Abstract

Grain boundary wetting as a preliminary stage for zinc induced grain boundary weakening and embrittlement in a Zn coated press hardened 20MnB8 steel was analyzed by means of electron backscatter diffraction, Auger electron spectroscopy, energy dispersive X-ray analysis, atom probe tomography and transmission electron microscopy on the nanometer scale. Microcracks at prior austenite grain boundaries were observed and structures that developed after microcrack formation were identified. Zn/Fe intermetallic phases with grain sizes smaller than 100 nm in diameter are present at the crack surfaces and at the wedge-shaped crack tips. In order to get a complete picture, including the microstructure before cracking, an undeformed, electrolytically coated reference sample which underwent the same heat treatment as the press hardened material was investigated. Here, Zn, in the order of one atomic layer or less, could be found along prior austenite grain boundaries several micrometers away from the actual Zn/Fe phases in the coating. Such a grain boundary weakening by Zn wetting of prior austenitic grain boundaries during austenitization and/or hot forming is a necessary condition for microcrack formation.

# 1. Introduction

Press hardened ultra-high strength steel (PHS) is the material of choice for structural components in lightweight automotive body technology to reduce vehicle weight and increase passenger safety [1]. PHS blanks are heated up to temperatures above 870°C and subsequently quenched and hardened. In the indirect process, preformed parts are heated up and quenched in a die. In the direct process, a flat blank is heated up and forming as well as quenching is done in one step. The steel surface is protected by a zinc (Zn) coating from oxidation during the final annealing process, in which the former ferritic and bainitic structure transforms again to austenite (γ-Fe). During the subsequent quenching, the steel sheet is formed by hot stamping leading to a hard, martensitic structure. The hot-dip galvannealed Zn coating does not only prevent the steel from oxidation during annealing but also performs excellently as corrosion protection [2].

However, Zn coatings on steel combined with heat treatments also present a challenge. Zn can embrittle the base steel and macroscopic cracks can occur during the further processing steps. Grain boundary (GB) weakening by Zn is supposed to be the underlying mechanism of this metal induced embrittlement (MIE), which is therefore highly influenced by different alloy compositions and their resulting microstructures. Already a monolayer or even a sub-monolayer of Zn at a GB has an impact on its strength [3]. The transport of the Zn to the GBs can happen by solid state diffusion (solid metal induced embrittlement, SMIE) [4], by liquid Zn (liquid metal induced embrittlement, LMIE) or even due to Zn vapor [5]. Since Zn liquifies at quite low temperature above 420°C, LMIE is the most prominent process and has been a topic for more than 100 years and is still in focus of recent research [6, 7, 8, 9, 10, 11, 12, 13, 14].

Potential counteractions are the binding of Zn by pre-cooling in Fe-Zn phases such as the $\Gamma$-$Fe_4Zn_9$ phase with a melting point $T_m$=782°C or by oxygen [15, 16, 17] in ZnO ($T_m$=1975°C, see Table 1). MIE is well under control in the manufacturing process of PHS-systems by these effects and measurements described in the review article from Ling et al. [18]. For example, by adding elements like boron to the underlying steel grade, one can lower the conversion into softer microstructures. Thus, it is possible to use less constrained cooling curves. However, the models of the different MIE mechanism are still under debate. Understanding of these mechanisms is a crucial step for the further development of the PHS technology and of new Zn-coated ultra-high strength steel grades in general. In recent works [8-13], steel samples were tensile stressed at high temperatures so that liquid Zn could flow into the new appearing cracks and form intermetallic Zn-Fe phases afterwards. We are convinced that the formation of these phases hides the view on the initial stages of embrittlement, where the first interaction between the liquid Zn and the base steel occurs. This can prevent a better understanding of the underlying mechanisms. Thus, the aim of this study is to focus on the first wetting of GBs with Zn on the nanometer scale. Therefore, we use samples, which underwent a heat treatment similar to the temperature history of the industrially manufactured ones. In this way, we can simulate possible production failures and investigate the initial stages of grain boundary weakening.

*Table 1. possible phases at the surface of the PHS-steel*

| Phase | composition | Lattice constant [Å] | Crystal system | Space group | Space group number | Melting temp. [°C] |
|---|---|---|---|---|---|---|
| ferrite | Fe | a = 2.8710 | Cubic | I m -3 m | 229 | - |
| Γ | $Fe_4Zn_9$ | a = 8.9820 [19] | Cubic | I -4 3 m | 217 | 782 |
| $Γ_1$ | $Fe_{11}Zn_{40}$ | a = 17.9630 [20] | Cubic | F -4 3 m | 216 | - |
| Δ | $FeZn_{11}$ | a = 12.7870<br>c = 57.2220 [21] | Hexagonal | P 63/m m c | 194 | 670 |
| Z | $FeZn_{13}$ | a = 10.8618<br>b = 7.6080<br>c = 5.0610<br>β = 100.542° [22] | Monoclinic | C 1 2/m 1 | 12 | 530 |
| Zinc | Zn | a = 2.6640<br>b = 4.9469 | Hexagonal | P 63/m m c | 194 | 420 |
| zincite | ZnO | a = 3.2489<br>b = 5.2049 | Hexagonal | P 63/m m c | 186 | 1975 |

# 2 Experimental Details

## 2.1 Material and Sample Processing

For this study, a galvanized transition retarded boron steel (20MnB8 + GA90/90) was used, which is distributed by voestalpine Stahl GmbH under the brand name phs-directform 1500. Alloy elements in the steel are: C 0.2 wt%, Si 0.2 wt%, Mn 2 wt%, 30 ppm B. The thickness of the steel sheets is approximately 1.5 mm. The steel is hot-dip galvanized in a Zn-bath at 460°C containing 0.15 wt% Al and galvannealed with a final layer thickness of about 12.5 µm. A Fe-Zn phase diagram adapted for boron steel can be found in Ref. [23] for a 22MnB5. This phase diagram represents also our system of a 20MnB8 better than the standard Fe-Zn phase diagram.

To simulate MIE cracks, we deformed sample strips with sizes around 20x120 mm in a die [17]. First, the strip of the Zn-coated base material was heated in a radiation furnace to 900°C and annealed for 45 seconds at a temperature above 870°C. The sample was subsequently cooled in air to 600°C and finally bent in the tool reaching a temperature of 100°C as presented in the thermal processing curve in Fig. 1a. This precooling down to 600°C corresponds to the industrial process preventing so-called 1$^{st}$ order LMIE due to liquid Zn. Cracks appeared at the stretched coating-steel surface, as indicated in Fig.1b, and continue into the base steel, as depicted in Fig. 1c, a SEM cross section image.

The interaction of the Zn-coating with the base steel was analyzed on a second set of samples. The large interface roughness between the base steel and the Zn-coating of the steel substrate offered some challenges regarding the preparation of samples for transmission electron microscopy (TEM) by focused ion beam (FIB). In addition, the large Zn-coating thickness and ferrite grain sizes compared to the final specimen size achievable by FIB milling made a targeted preparation complex and lead to many failed attempts. To avoid these challenges, we fabricated well-defined test samples, which minimizes the impact of production artifacts on the industrial sample surfaces (for instance laps or double skin from rolling). We used a similar galvannealed 20MnB8 steel to ensure the same microstructure and removed the Zn-coating via polishing to get a clean steel surface. Subsequently, a new Zn layer was deposited on the steel sample by electroplating creating a well-defined steel substrate-Zn interface. This strip was then annealed in the same way as the first type of samples for 45 s above 870°C and remained undeformed. It should be noted that this substrate-Zn interface does not contain a $Fe_2Al_5$ inhibition layer, which excludes a potential influence of this layer on GB wetting. However, such an inhibition layer is not present in the hot-dip galvanized samples, since any present inhibition layer is dissolved quite early in the galvannealing process.

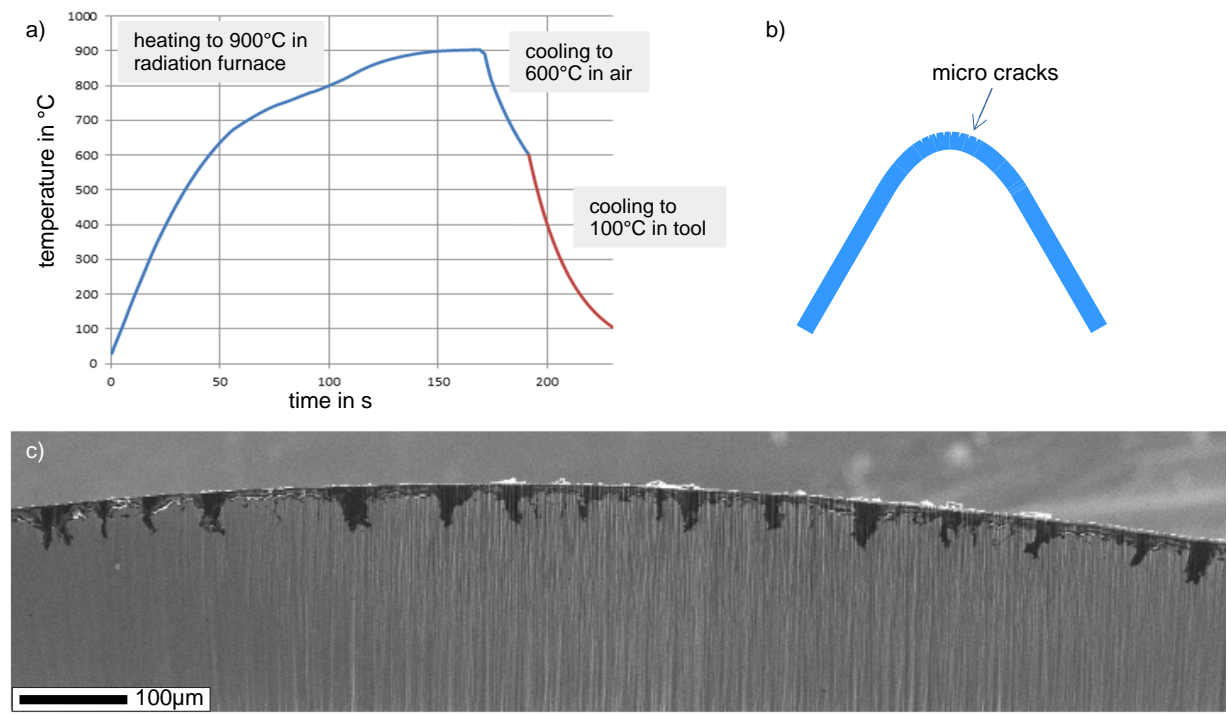

Fig. 1. a) Thermal processing curve of the PHS-sample (hot-dip galvanized 20MnB8) in order to simulate realistic production conditions and provoke microcracking, b) sketch of the deformed sample with microcracks at the stretched surface, and c) secondary electron image of micro-cracks (filled with epoxy resin to avoid redeposition during sample preparation) at the bent surface in a cross section view.

## 2.2 Analytical Tools and Measurement Conditions

Cross sections were analyzed in a Zeiss Supra 35 field emission scanning electron microscope (SEM) with an X-Max$^N$ 80 mm² energy dispersive X-ray detector (EDX) and a NordlysNano electron backscatter diffraction (EBSD) detector. Both detectors are from Oxford Instruments and can be used simultaneously. The cross sections were produced via argon ion sputtering in a Leica TIC 3X at -100°C with an ion beam energy of 8keV. The sample surface was protected from re-deposition in the sputter chamber with an ink layer of a black felt tip pen and a copper tape.

The crystal structures and lattice parameters were identified by XRD (X'Pert PRO MPD, Malvern Panalytical and Rietveld refined, see Table 1). The found phases were transferred into the Aztec® program and were used as reference list for phase identification during EBSD mapping. The phases with the highest volume fraction, namely bcc-iron (ferrite), Γ- and Γ$_1$-phases are cubic and can hardly be distinguished. Also, the hexagonal phases Zn, ZnO and δ are too similar to be distinguishable with EBSD. Additionally, the $Γ_1$ lattice constant is exactly the double of the Γ phase, which complicates also the identification of these phases by XRD.

Cross sections were also analyzed with Auger electron spectroscopy (AES) in a JAMP 9500F from JEOL. The instrument is equipped with a fracture stage that is flange-mounted to the main chamber. This allows breaking samples after cooling with liquid nitrogen and measuring the fractured surfaces without leaving the ultra-high vacuum (base pressure $5·10^{-8}$ Pa). All Auger spectra were taken in constant retard ratio mode (CRR) with an energy resolution of about ΔE/E = 0.5%. The primary electrons were accelerated to 30keV at currents between 10 and 20nA. Auger element mappings were conducted in constant analyzer energy mode (CAE). The depth information is about 6 nm and the lateral resolution 10 nm. Although the instrument is equipped with an Ar ion sputter gun, the sample surfaces were not sputtered, in order to keep the original preparation state.

Transmission electron microscopy (TEM) was performed in a JEOL JEM-2200FS instrument, equipped with an Oxford Instruments X-Max$^N$ 80T EDX-detector, at 200 keV primary beam energy. The instrument was operated in Scanning TEM mode, using a bright field (BF) and a high angle annular dark field (HAADF) detector. Sample preparation for TEM was conducted by focused ion beam (FIB) cutting in a Zeiss 1540 XB SEM. The surface of the sample with microcracks was covered prior to FIB cutting with epoxy resin M-Bond 610 from Vishay Micro-Measurements in order fill the crack and potential pores to avoid re-deposition of the sputtered material.

Atom probe tomography (APT) was used to obtain a detailed chemical composition of the Zn coating and the base steel. Additionally, segregation at defects in the base steel was investigated via this method. FIB milling using a FEI Helios Nanolab 600i was employed to prepare specimens for APT. Specimens were prepared employing the standard liftout procedure. Some more details about the liftout procedure are given in the supplementary information section. APT tips were sharpened at 30 keV by annular milling. A final low-energy milling step at 5 keV for 1 min was applied to reduce Ga contamination at the tip surface. APT experiments were conducted in a Cameca LEAP 5000X HR and a LEAP 5000 XS, both operated in laser mode. Pulse frequencies of 333 kHz - 625 kHz and 125 kHz - 333 kHz were used together with detection rates of 1.5 % - 3 % and 1.3 % - 1.7 % for the XS and the HR device, respectively. Temperatures between 40 K and 60 K and laser energies between 35 pJ and 60 pJ were used for both devices. In order to cross check the composition obtained by APT in laser mode, two additional measurements were performed in voltage mode at a temperature of 50 K, detection rate of 1.2%, a pulse fraction of 15% and a pulse frequency of 250 KHz. The compositions were found to be the same with both operation modes of the atom probe instrument. The commercial software IVAS (version 3.8.4) was used to reconstruct the APT specimens.

# 3 Results
## 3.1 Metal induced embrittlement cracks

The first analyzed structure was taken from the hot dip galvanized PHS strip, which was deformed during cooling. The cross section of the sample was subsequently investigated via EBSD and EDX in the SEM Zeiss supra35 and the results of this measurement is presented in Fig. 2. The general structure of the coating system is indicated in the band contrast (BC) image of Fig. 2a. Starting with the base steel, it is i) martensite, ii) Zn-saturated ferrite (α-Fe(Zn)), iii) Zn/Fe-phases (mostly Γ-phase), iv) ZnO (zincite if crystalline). In the band slope (BS) image (Fig. 2b), martensite can be clearly distinguished from ferrite. The dark regions correspond to martensite, due to the high density of sub-grain boundaries and dislocations. Fig. 2c shows the EBSD phase map, in which the three phases bcc-Fe (martensite and α-Fe(Zn)), Γ-phase and ZnO are shown. Close to the α-Fe(Zn), the Γ-phase was partially incorrectly identified as bcc-iron phase, because the crystal structures of both phases are too similar. Another feature, that is well visible as dark areas in the Γ-phase region, are pores reaching from the surface far into the Γ-phase. Fig. 2d combines the BC with the EDX Zn $K_\alpha$ and Fe $K_\alpha$ maps. A sharp boundary is visible in α-Fe(Zn) grains that touch the bulk martensite, where the Zn concentration decreases from about 30 wt% to approximately 0 wt%. This leads to a small region of Zn-free ferrite (α-Fe) grain sections separating the Zn in the α-Fe(Zn) grains from the bulk martensite. Similar findings of an additional phase separation the α-Fe(Zn) from the bulk martensite at the interface region was found in related Zn-coated steel systems [14, 24] and is discussed further below. The measured chemical concentration in the Zn-rich part of the α-Fe(Zn) is in wt%: Fe 69.5, Zn 29.8, Al 0.2, Mn 0.2, Si 0.1, Cr 0.1. In the underlying martensite the following concentrations in wt% were found: Fe 97.6, Zn 0.2, Al 0.2, Mn 1.4, Si 0.2, Cr 0.3. No extended Zn-phases were observed in the crack in the martensite region.

The surrounding area of a crack tip (see Fig. 3) was analyzed with the same techniques as the survey mapping, but at higher magnification. The fine martensitic structure around the crack can be seen in Fig.3a. The Zn $K_\alpha$ EDX map shown in Fig.3b reveals that the crack surface is covered with Zn. Furthermore, the Zn concentration at the two marked crack tips is higher compared to the remaining cracked surface. Fig. 3a shows also some misorientation angles from opposite sides of the crack tips. The austenite grains that are present at higher temperatures, transform into martensite during rapid cooling. The transformation follows the Kurdjumov-Sachs mechanism, which predicts the orientation relation of martensite grains to its prior austenite grain. No possible grain orientation between 21.06° and 47.11° does exist according to this relation [25, 26] for martensite grains originating from the same austenite grain. Thus, martensite grains exhibiting an angle between these values have to originate from different austenite grains. And indeed, the measured misorientation angles of grains from opposite sides of the crack tip between 24° and 37° indicate cracking along prior austenite grain boundaries (PAGBs). This is in good agreement with recent results by Takashi et al. [7].

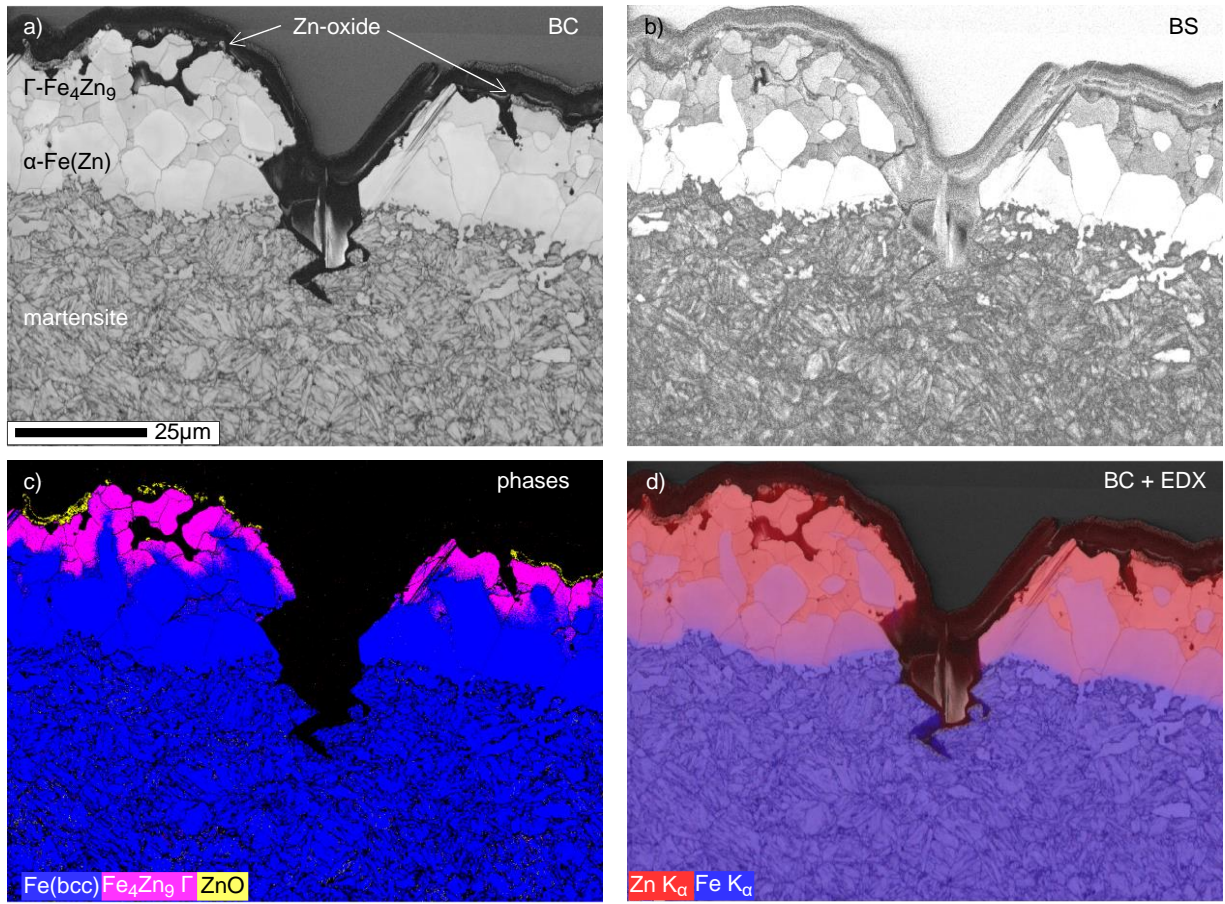

Fig. 2. Cross section of a crack in the hot-dip galvanized, deformed sample: a) EBSD band contrast; b) band slope contrast, c) overlay of ferrite, Γ-phase and zincite and d) an overlay of band contrast and EDX Zn $K_α$ and Fe $K_α$

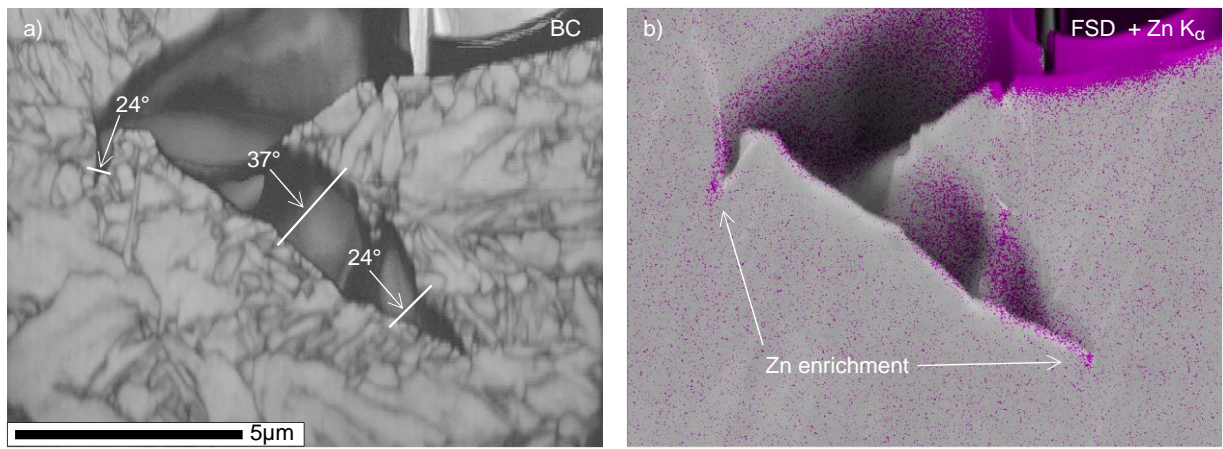

Fig. 3. a) EBSD band contrast mapping of the crack tip and b) front scatter detector image overlaid with EDX Zn $K_α$ signal.

To get a deeper insight into the composition of the crack tip, Auger measurements were performed (see Fig. 4). Since EBSD mapping leads to carbon contamination in the map area and Auger electron spectroscopy is extremely surface sensitive, a new crack similar to that one shown in Fig. 3 was chosen for AES measurements. Fig. 4a shows the secondary electron image of a similar crack tip from the same sample, where six Auger spectra (Fig. 4d) were taken at the marked positions. Oxides and carbon contamination form rapidly during sample handling after preparation in ambient atmosphere. Therefore, traces of C and O are present in all spectra, which is not further discussed on the assumption that this does not affect the distribution of the other elements. On the contrary, oxidation can even help to reduce the mobility of Zn and to fix the original Zn-distribution. The only two other elements which can be observed in the spectra are Fe and Zn. First, this is a proof that the sample preparation is suitable, because no re-depositions of elements like Cr, Ni, Cu or Al from the stainless steel walls of the sputter chamber or from Al- and Cu-tapes used to protect the sample is present. Fe is found in all spectra, while Zn only occurs at the crack surface (spectra 4-6) and at the crack tip (spectrum 1). No Zn can be found in spectra 2 and 3, which indicates the absence of significant Zn bulk diffusion. The Zn concentrations in spectra 4-6 are comparably low, deviations could be explained by the different angles of the surfaces facing the detector. The highest Zn concentration is found in spectrum 1 at the crack tip. A quantification of this spectrum gives the following concentration in at%: C 45.5, O 33.8, Fe 14.3, Zn 6.4. The Fe/Zn ratio of ~7/3 indicates that the material at this position was a Zn-Fe phase with 30 at% Zn before oxidation. Furthermore, the Auger mapping (Fig 4b and 4c) using the $Zn_{LMM}$ signal reveals a triangular, pointy shape of the Zn-enriched area at the crack tip along the crack direction.

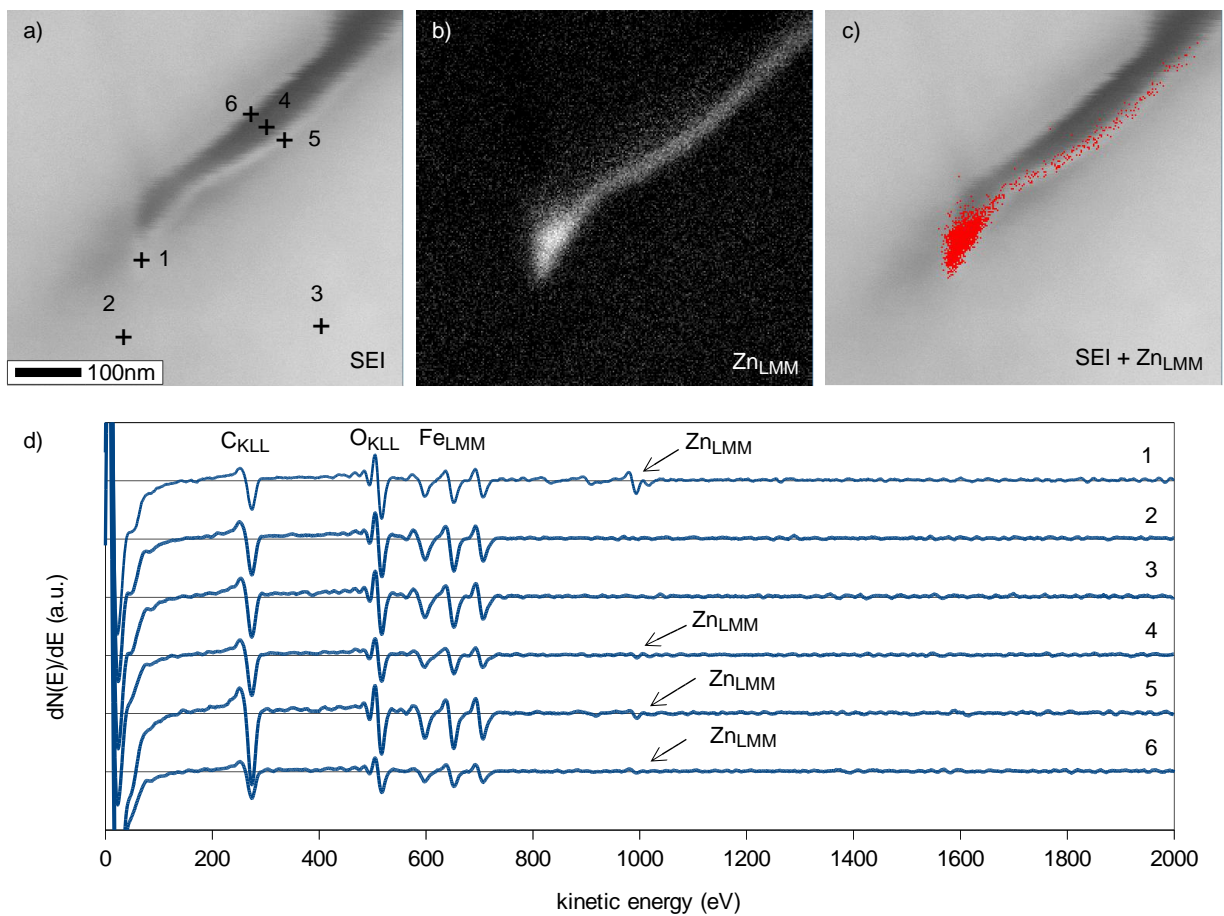

Fig. 4. a) Secondary electron image of a crack tip in the hot-dip galvanized, deformed sample, prepared with cross section argon sputtering, b) corresponding $Zn_{LMM}$ Auger mapping, c) the overlay of both and d) Auger derivative spectra. The surface of the crack is covered with Zn. The crack tip was a α-Fe(Zn) wedge smaller than 100nm before oxidation.

An additional crack of the analyzed sample was prepared as a FIB-lamella for STEM investigations. Fig. 5a shows the HAADF image of the analyzed area. The dark area in the top right corner is the crack, filled with epoxy resin to avoid re-deposition onto the surface during FIB-cutting. The vertical strips are thickness variations of the lamella, a well-known FIB-preparation artefact called curtaining effect, resulting from the inhomogeneous surface and grain structure of the complex steel sample. The lamella was subsequently analyzed with EDX. Fig. 5b shows the Zn $K_α$ map, where three different aspects are evident. It indicates three effects. First, it confirms the observation made in Fig. 4d that Zn is present at the crack surface. Second, it also shows a triangular shaped Zn phase at the crack tip similar to Fig. 4b and c. Third, a new finding is that Zn penetrates further into the steel at the tip of the Zn wedge. The position of the Zn at the edge of the crack is confirmed in Fig. 5c, which displays the overlay of the HAADF STEM image with the Zn $K_\alpha$ map. Additionally, a further structure found in this EDX measurement is displayed, small precipitates containing the alloying element titanium.

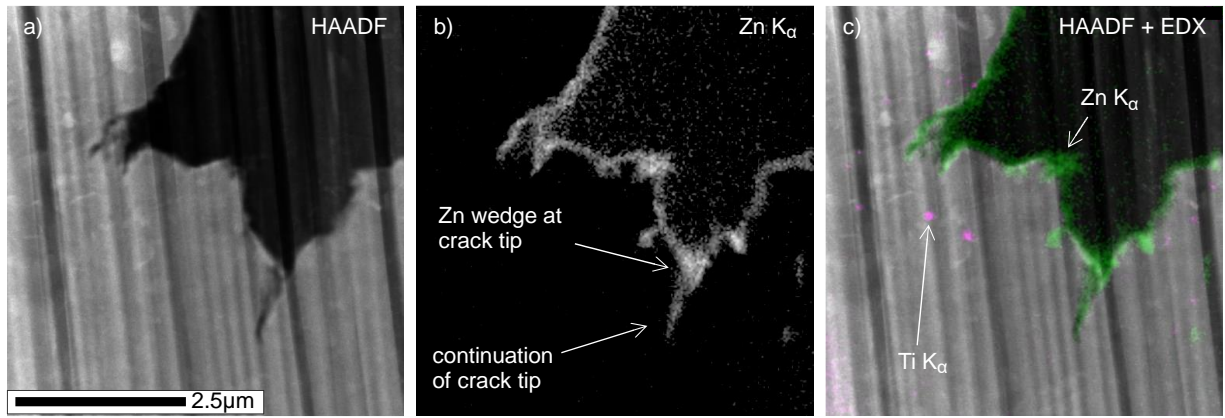

*Fig. 5. a) HAADF STEM image of a crack tip in the hot-dip galvanized, deformed sample. b) Zn EDX mapping made in TEM and c) overlay of both. The Zn distribution shows, that the surface of the crack is covered by Zn. It also shows the Zn wedge at the crack tip. There is also a continuation of the Zn into the base material, observable at the tip of the Zn wedge.*

## 3.2 Interaction of Zn with the undeformed base steel

An important question is how far Zn penetrates into the base steel before cracking, and whether the penetration of Zn into the GBs requires bending of the sample and an external force during cooling. For this purpose, the second experiment with the Zn-electroplated, undeformed base material was performed, which was annealed in the same way as the MIE affected sample.

The cross section of the sample was polished and subsequently measured via EBSD and EDX similar to the first, deformed sample. Fig. 6a shows the band contrast image of a sample surface cross section. Overlays of this band contrast with EDX data and EBSD phases are shown in Fig. 6b-d. As expected, and proofing our experimental concept, we found similar structures as in the bent sample shown in Fig. 2. The bulk consists of martensite. The next layer is composed of α-Fe(Zn). Here again, a sharp internal boundary occurs where the Zn concentration increases from approximately 0 to 30 wt% leading to a thin Zn-free region within the α-Fe grains located at the α-Fe(Zn)/martensite interface. This Zn-free ferrite region is an almost continuous layer across the entire interface, interrupted only by some regions where the Zn-containing α-Fe(Zn) grains touch the martensitic base steel directly (see Fig. 6 and Fig. S2 in the supplementary section). The next layer consists of α-Fe(Zn) surrounded by Zn/Fe-phases (Γ- and δ-phase). The top layer is a continuous crystalline ZnO film with large pores beneath. This ZnO layer is thicker for the unbent sample in contrast to the sparse occurrence of ZnO on top of the bent sample. The reason is that the Zn layer of the bent sample is created by hot-dip galvanizing and galvannealing, in which the liquid Zn bath contains approximately 0.15 wt% Al. This leads to an additional $Al_2O_3$ layer on top of the coating, which reduces Zn oxidation [27]. This missing $Al_2O_3$ of the electroplated sample displayed in Fig. 6 leads to the formation of the thicker ZnO layer. Within the sensitivity of the combined EDX/EBSD measurement, no Zn is detected below the internal α-Fe(Zn) zinc boundary (see Fig. 6 and Fig. S2 in the supplementary section).

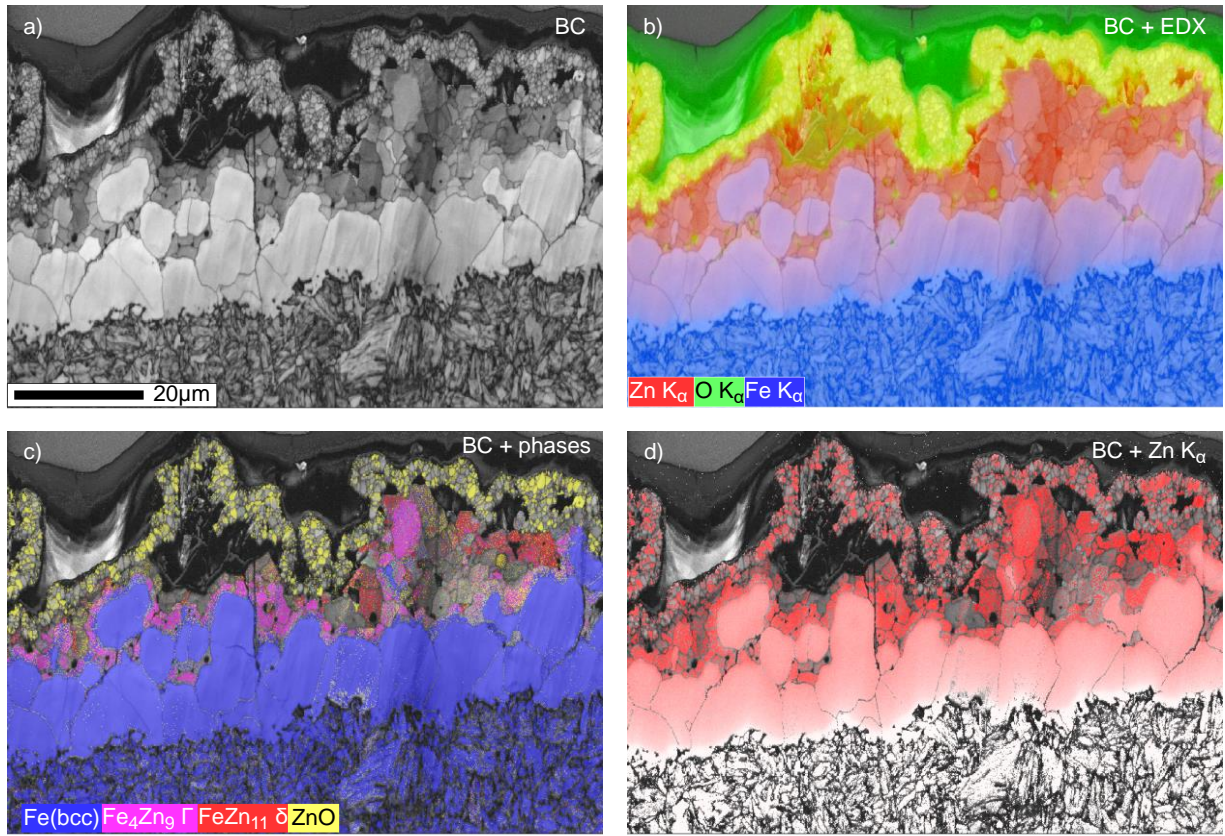

*Fig. 6. Cross section of a electrogalvanized, undeformed sample: a) EBSD band contrast; b) overlay of band contrast with EDX Zn $K_\alpha$, O $K_\alpha$ and Fe $K_\alpha$; c) overlay of band contrast with ferrite, Γ-phase, δ-phase and zincite; d) overlay of band contrast with EDX Zn $K_\alpha$.*

To obtain a detailed elemental composition, APT measurements were performed. As explained in more detail in the supplementary section (see Fig. S1-S3), multiple APT tips were measured along an approximately 10 μm long so-called liftout. In this way we generated the composition profile from the Γ-phase and the α-Fe(Zn) to the martensitic base steel shown in Fig. 7. The composition of the Γ-phase is 31.9 at% Fe, 68.1 at% Zn and a small amount of 0.033 at% Mn, which is very close to the stoichiometric composition of $Fe_4Zn_9$. A clear gradient in Zn composition with decreasing Zn content towards the base steel is present in the region of the α-Fe(Zn) phase. The Zn composition at the exact interface to martensite cannot be determined here, since the composition profile consists of discrete APT measurement points being approximately 2 μm apart. The graph shows differently steep increasing concentrations of Si, Mn, Al, Cr as well as C towards the base steel. The exact compositions of each APT measurement can be found in Table S1 in the supplementary section. The martensitic base steel exhibits a composition of approximately 0.4 at% Si, 1at% Mn, 0.03-0.048 at% Al, 0.18 at% Cr and 1.45-1.67 at% C. The Zn composition was found to be extremely low at only 0.002-0.003 at% (i.e. 20-30 ppm). The C content is higher than the nominal composition of 1 at% which could be an enrichment of C during the α-Fe(Zn) formation on the steel base side due to a low solubility of the formed phase. Other reasons could be local variations in C concentration, as well as C contaminations during any of the preparation steps.

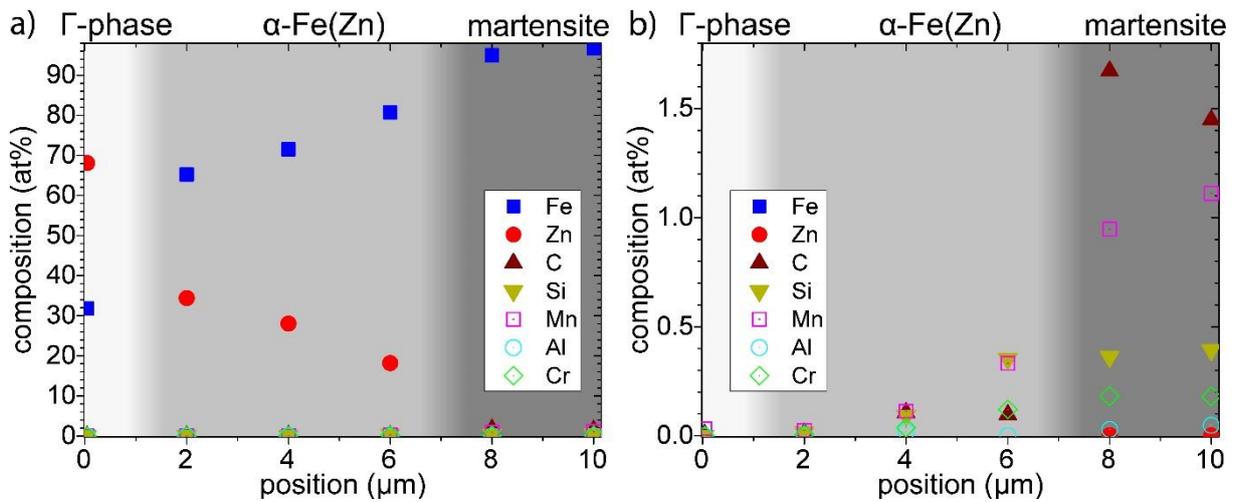

*Fig. 7. Elemental compositions determined by APT at the electrogalvanized, undeformed sample: a) shows the chemical composition of the Zn coating system and the base steel, ranging from the Γ-phase over the α-Fe(Zn) to the martensitic base steel (left to right). Each set of datapoints at the different positions represents an individual APT measurement. More details can be found in the supplementary materials (Fig. S1-S3). b) presents a magnified view of the low concentration regime from 0-1.8 at%. The graph shows the data from liftout number one (see supplementary Fig. S1 and S3 for reference). The exact values can be found in table S1 in the supplementary section.*

If this sample contains PAGBs weakened by Zn, they should open up preferentially during fracture [3]. Thus, we look for the presence of Zn in the martensite base steel below the α-Fe(Zn) on fractured surfaces prepared with a fracture stage, which is attached to the Auger instrument. This allows us to analyze the fractured surface *in situ* via Auger electron spectroscopy without leaving the ultra-high vacuum. In this way, we can avoid any surface contamination or oxidation. Fig. 8a shows such a fractured surface, where six Auger spectra measurement positions are marked. The image is also overlaid with an EDX mapping of the fractured surface that was performed after the Auger measurements. The EDX map reveals the interface between α-Fe(Zn) and martensite. The positions of the measured point spectra are all positioned beneath this interface. The only small regions with Zn in the martensite are fracture splinters and were avoided for the spectra measurements. Fig. 8b shows the Auger spectra, which were taken around the Zn energy range. A significant $Zn_{LMM}$ peak was found in spectra 1-4, which means that Zn is still present on the fractured surface 25μm away from the α-Fe(Zn)/martensite interface in the martensitic steel region. The Zn concentration of spectrum 4 was quantified by taking a complete spectrum from 0 to 2000eV (not shown here), which resulted in a Zn concentration of approximately 0.6 at% Zn. This Zn concentration found by AES on the fracture surface is much higher than the Zn content of the martensitic matrix determined by APT to be 0.002 at%. These results show that the fracture surface exhibits a local enrichment in Zn i.e. a Zn enriched GB that was weakened and therefore opened up preferentially upon fracture as intended with this experiment. The Zn concentration determined by AES is much lower than Zn concentrations measured at ferrite grain boundaries by Allegra et al. [28] after long time diffusion experiments. However, if the information depth of AES is estimated to be a few nm, the AES result of 0.6 at% Zn can be an amount of a Zn monolayer or a comparably thick Zn-Fe phase at the fractured surface. This is the same order of magnitude of the unwanted element as in the studies of Luo et al. [29] on a simpler system, in which a bilayer of Bi was found responsible for LMIE of austenite grain boundaries in Ni. For SEM-EDX it is impossible to determine such low Zn concentrations at surfaces, because the information depth is several hundreds of nm. And, indeed, no Zn was detected (besides the splinters that formed during the *in situ* fracture in the AES chamber) in the martensite regions with SEM-EDX on the fractured surface.

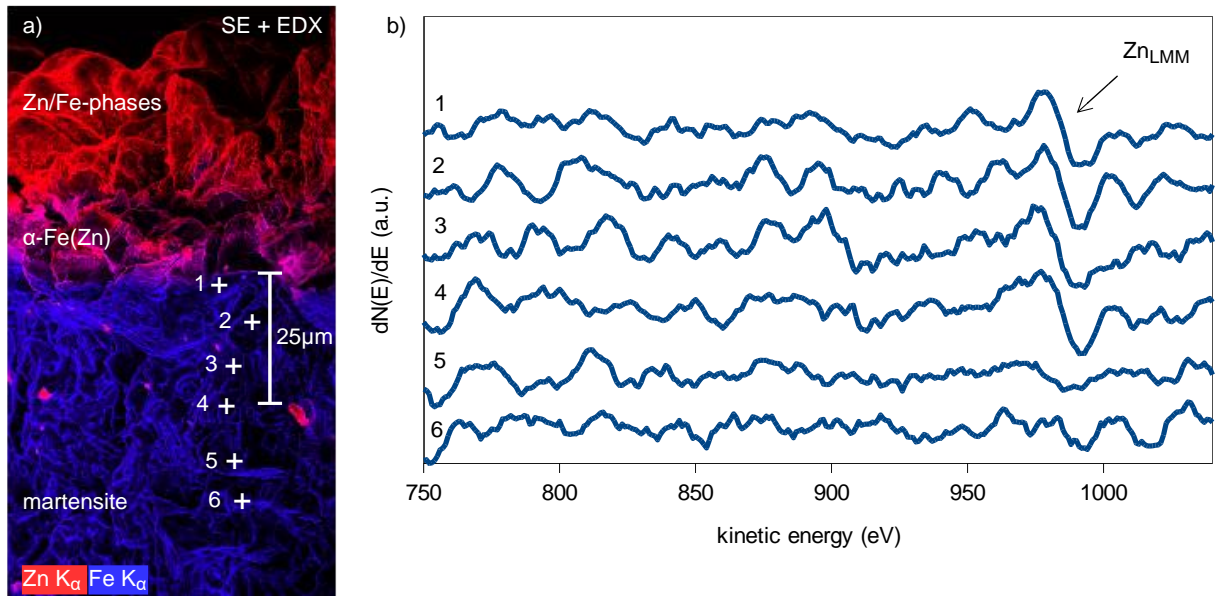

*Fig. 8. Six Auger measurement positions at a fractured surface of the electrogalvanized, undeformed sample prepared in a UHV fracture stage mounted to the Auger instrument. a) shows the overlay of the secondary electron image with the EDX Zn $K_α$ and Fe $K_α$ mappings. The corresponding differentiated Auger spectra are displayed in b). One can see, that Zn can be detected down to 25μm below the Zn boundary in the α-Fe(Zn) layer. Subsequently the area was EDX mapped in an SEM to ensure the correct Auger measurement positions below the α-Fe(Zn).*

These low amounts of Zn make it also quite difficult to identify a Zn-wetted GB and prepare it for further measurements like APT, which would be perfectly suited to determine such small concentrations. Our APT measurements were an attempt preparing Zn-wetted PAGBs within APT tips. For the liftouts, we target grain boundaries in the α-Fe(Zn) that continued straight into the martensite and where no Zn-free ferrite section was present. Unfortunately, we could not unambiguously identify a PAGB in the APT measurements, for details see the supplementary section. In General, we think that usually not all PAGBs are covered with Zn and due to the difficult pre-identification of potential Zn-wetted GBs, there is only a low probability of finding a Zn-wetted PAGB within APT tips with a limited number of attempts. A slightly higher probability can be reached with TEM, because compared to APT tips a larger volume can be sampled by using large FIB-lamellae. Depending on the sample thickness and due to the higher acceleration voltage, the interaction volumes in TEM-EDX is only several hundreds of $nm^3$ large increasing the sensitivity to identify the Zn-wetted PAGB and in general the lateral resolution compared to SEM experiments.

We successful prepared a lamella from the undeformed electroplated sample in order to perform an EDX-STEM analysis with higher sensitivity. Fig. 9a shows a bright field STEM image of a prior austenite grain. On top of it lies the α-Fe(Zn) layer identifiable by a more homogeneous light gray color. Below, the lath-like structure of martensite is visible. Its high dislocation density, as well as internal stress leads to a more complex appearance. One can recognize two GBs enclosing a group of martensite grains, which, according to our interpretation, represent a prior austenite grain. The GBs are marked by two EDX mapping areas and arrows pointing toward them. The GB displayed in Fig. 9b was confirmed to be a PAGB via transmission Kikuchi diffraction. The misorientation between the two neighboring grains was measured to be approximately 5°, which is clearly between the two possible orientations according to Kurdjumov-Sachs, 0° and 10.53° [25, 26]. The EDX-mappings were recorded over several hours in order to have sufficient counts for a reliable signal evaluation of elements with low concentration. Fig. 9b exhibits a Zn enrichment along the PAGB deep into the martensite, where Zn penetrates into the base steel at least 3µm. Following the PAGB, no Zn could be measured any more after a few µm. Either the coverage of the GB decreases really below the detection limit, or it is a geometrical effect. The PAGB plane could be oriented in such a way that it is not parallel to the e-beam anymore, which would lower the small signal to the noise level.

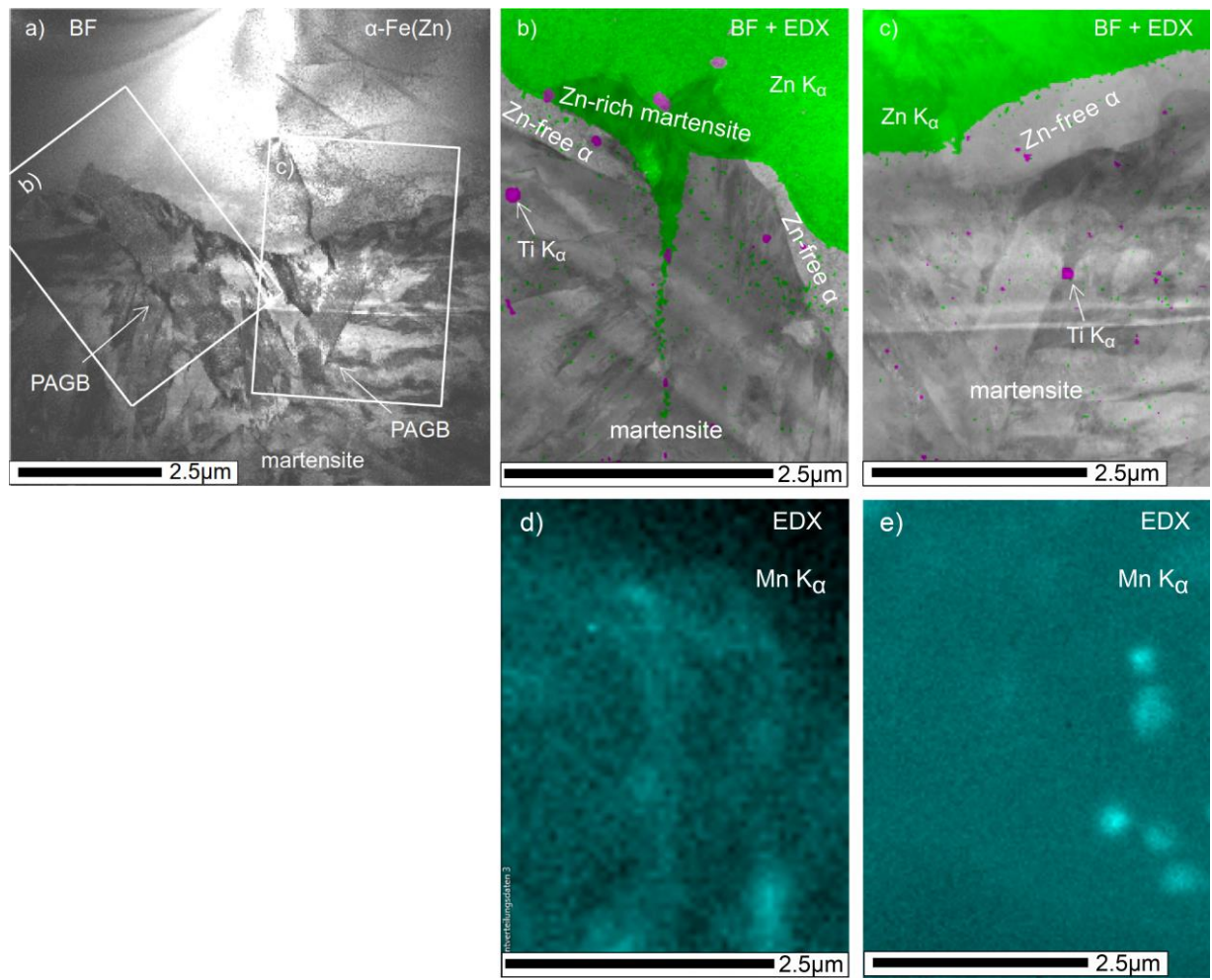

*Fig. 9. STEM measurements at the electrogalvanized, undeformed sample. a) shows a STEM bright field image of a prior austenite grain in the martensite below the α-Fe(Zn). b) depicts the corresponding EDX signal of Zn $K_\alpha$ (green) and Ti $K_\alpha$ (purple) in an overlay with the bright field image, which shows the α-Fe(Zn) at the top. One can also see Zn along the prior austenite grain boundary. c) shows the EDX mapping for the second, neighboring austenite grain boundary, where no Zn wetting was observed. d) and e) are the corresponding Mn $K_\alpha$ maps for the two PAGBs.*

Interestingly, this specific PAGB which is clearly Zn-covered is directly touching the Zn-containing region of the α-Fe(Zn) i.e. no Zn-free ferrite is present. In this region, the Zn-free lower part of the α-Fe(Zn) is missing and replaced by two small Zn-containing martensite grains. Another observation can be made in the EDX mapping of the second PAGB displayed in Fig. 9c. At this PAGB no Zn could be found and no diffusion of Zn along the PAGB is present, instead a continuous layer of Zn-depleted ferrite separates the Zn-saturated part from martensite base material. In this region, the Zn concentration decreases abruptly from 30 wt% to approximately 0 wt% within the α-Fe(Zn) grain. This observation matches the results from SEM-EDX and EBSD seen in Fig. 2 and 6. For comprehensiveness, also the Ti signal is shown in these maps, which was the second alloy element we found to have a distinct distribution and forms different precipitates in the PAGBs surrounding. Fig. 9d and e display the corresponding Mn $K_\alpha$ signal. We found an increased Mn content in the surrounding of the Zn wetted PAGB. In the region of the second, non-Zn-wetted PAGA no Mn segregation correlated with the GB could be observed. At the same time small regions with significantly increased Mn content are present. These Mn enriched regions are presumably Mn rich precipitates such as Mn-Cr-carbides present in the 20MnB8 steel, which partially dissolve during the annealing step above 870°C. Additional measurements at different lamellae reveal similar results. We found in most cases at least one Zn-wetted grain boundary in about 10µm large lamellae, but not all PAGBs were wetted. This supports our argument that only some specific PAGBs are Zn-covered.

Different line scans are extracted from the mapping of Fig. 9 as indicated in Fig. 10a. Fig. 10b shows a line profile starting in the martensite and crossing the Zn-free ferrite into the α-Fe(Zn). It can be clearly seen that the Zn signal decrease abruptly at the interface α-Fe(Zn)/Zn-free ferrite to almost zero. The decrease is even more abrupt than shown, because the interface between the two phases is not perpendicular to the extracted line profile, thus the signal is smeared out due to geometrical reasons. For comparison, Fig. 10c is a line profile from the martensite through the Zn-containing martensite into the α-Fe(Zn). Here the Zn signal exhibits a small plateau of around 6at% Zn at the position of the Zn-containing martensite, which fully fits to a possible Zn-solubility in austenite during annealing according the phase diagram in ref. [23] (see Fig. 11). Both line profiles display additionally the Mn signal. Due to the general accuracy of EDX quantification, statements about the Mn are only possible to a limited extent. However, it seems that a small Mn increase is present at the interface martensite/Zn-free ferrite in Fig. 10b, after that the Mn decreases continuously which is also in accordance to the APT measurements (Fig. 7b). In Fig. 11c on the other hand, it seems that the Mn content is slightly increased in the Zn-containing martensite, which was at the annealing stage, Zn-containing austenite. This points to the importance of Mn, additionally to C, in the stabilization of the different phases (e.g. austenite during quenching) at the coating – steel interface. We summed up all spectra from horizontal lines in the marked rectangle of Fig. 10a to the quantitative line profile in Fig. 10d. This line profile reveals the extremely low concentration of Zn along the grain boundary. The increased Mn content at the PAGB is also displayed. Due to noise and the Cu $K_\beta$ (8.9keV) peak, which overlaps Zn $K_\alpha$, the statistical threshold for the quantification cannot be set so that the Zn concentration reaches zero far away of the GB. The increased Zn concentration at the GB is clearly visible and lies above the background. An estimation of the interaction volumes of AES and STEM, which are the same order of magnitude, shows that AES and STEM-EDX quantification values of 0.6 at% and 0.2at% fit well together. This strongly indicates the presence of a small amount of Zn or a Zn-Fe-phases at PAGBs. Wetting with nm or sub-nm thick layer or more complicated GB complexions several µm deep along the PAGBs are conceivable.

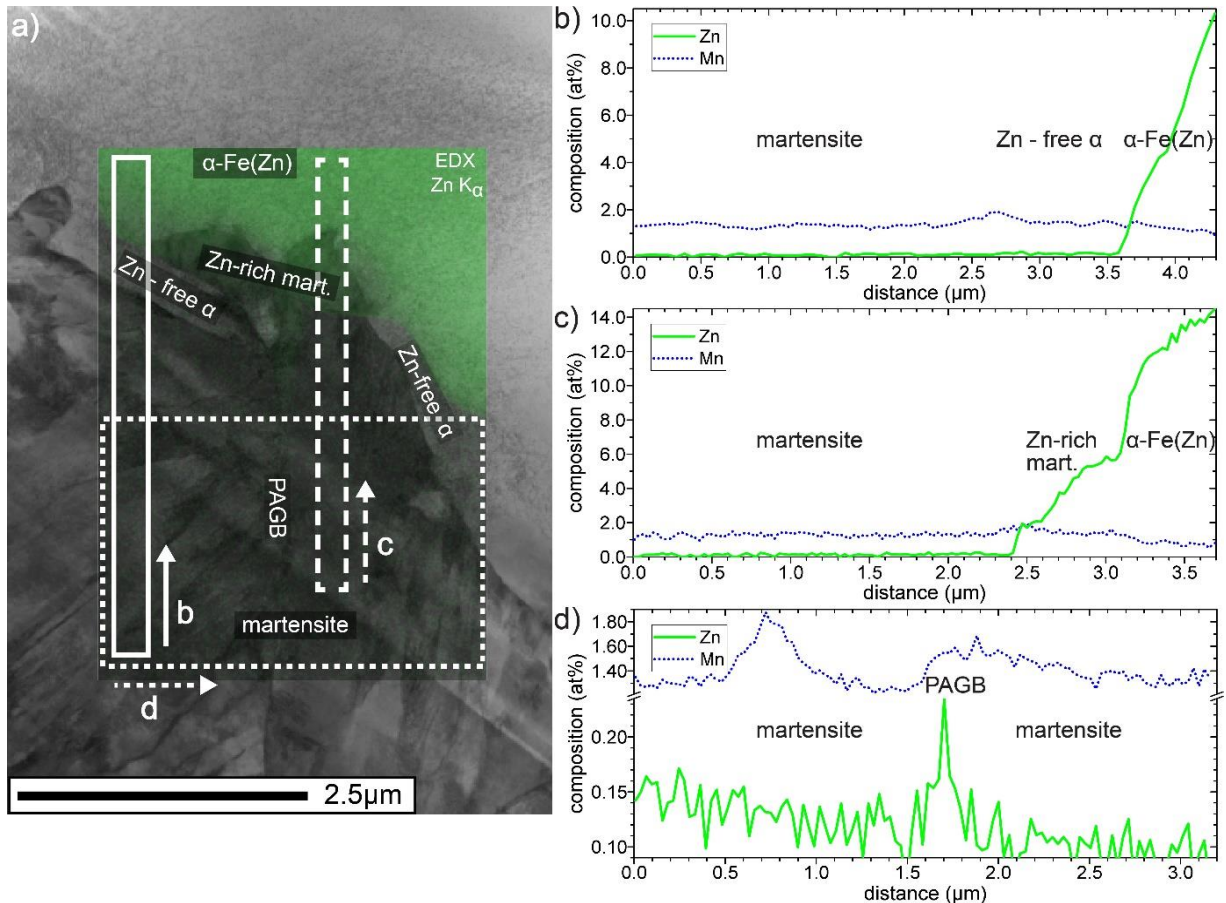

*Fig. 10. a) is the combined Zn K$_\alpha$ (green) and Fe K$_\alpha$ (red) mapping in an overlay with the bright field image of the wetted PAGB from the electrogalvanized, undeformed sample. b) displays the quantitative line profile of the corresponding rectangular region in Fig. 10a starting at the arrow and crosses the Zn-free ferrite. c) is the corresponding line profile through the Zn-containing martensite. d) shows a quantitative line profile obtained from the corresponding rectangular region in a). Here the line profile is generated from the sum spectra of all horizontal lines from left to right and display the concentration changes cutting through the PAGB.*

# 4. Discussion

Our investigations clearly prove the presence of small amounts of Zn at cracked surfaces and at crack tips. Zn induced crack formation at PAGBs is the reason for MIE in Zn-coated steel systems and different models are discussed in literature [9, 11, 14, 23]. Strategies to avoid MIE in modern PHS systems use these facts and follow strategies to binding high mobile liquid Zn or Zn vapor by oxidation to hampering the transport of Zn to the crack tip or to generally allow the Zn to form stable phases by extended annealing holding times [12, 13, 17, 23]. Investigations after the crack formation highlight the Zn distribution and crack propagation, but only allow a limited statement whether the found Zn-phases were originally responsible for cracking, or whether they were subsequently formed by Zn filling the crack. The question how the initial stage of the embrittlement, i.e. the weakening of the GB, occurred and thus the precondition of cracking is established, can hardly be answered on cracked samples. Thus, our undeformed samples allow a view on these first stages.

At typical austenitizing temperatures, in our case above 870°C, one can conclude from the B-steel-Zn phase diagram [23] (see Fig. 11) that the liquid Zn is enriched with Fe and that at the same time solid α-Fe(Zn) with up to 40 at% Zn can be formed at the contact region to the austenite (γ-Fe) matrix. Even if Zn can diffuse along austenite GBs leading to Zn-rich pockets, the α-Fe(Zn) formation is mainly governed by the liquid Zn γ-Fe reaction and solid state bulk diffusion. The reason is that the Zn diffusing fast along GBs can also be dissolved relatively fast in the bulk γ-Fe and forms α-Fe(Zn). Depending on the increased austenite region in the B-steel-Zn phase diagram, additionally the formation of Zn-rich austenite γ-Fe(Zn) can be possible [23, 24]. After quenching, this can lead to a region of Zn-rich martensite between the α-Fe(Zn) and the bulk martensite, which was reported in related Zn-coated systems [14, 24]. This Zn-rich martensite differs in its structure from the bulk martensite. Diverging element gradients due to different diffusion properties of Fe, Zn and the alloy elements as well as the different solubilities of the alloy elements in the α-Fe(Zn) and the austenite (with later transforms to martensite), lead locally to quite different alloy concentrations and a different transformation behavior compared to the bulk region. The work of Järvinen [24] shows that the formation of this Zn-rich martensite layer strongly depends on the C content of the steel grade. An increased C content in the steel leads to a more pronounced Zn-rich martensite layer.

Fig. 11. Calculated Fe-Mn-C-Zn-Cr-Si-Al phase diagram. Reprinted from reference [23] -> Please see Fig. 1 in [23].

Our 20MnB8 steel differs from the steels in ref. [24] by a lower C and a higher Mn concentration. It is probable that in our case, the diverging element gradients and thermal processing parameters lead to a small region attached α-Fe(Zn) that is free of Zn and can transform to the Zn-free ferrite described in the results section. In detail, the higher diffusion rates of the carbon compared to the Fe, Zn, Mn and the other alloy elements lead to a C-depleted region near the α-Fe(Zn), which can undergo the transformation to ferrite instead to martensite. Fig. 7b clearly shows the behaviour of diverging diffusion profiles. The concentration of the alloy elements in the martensitic region at position 8 and 10 in Fig. 7 is around 1.5at% C, 1at% Mn, 0.4at% Si and 0,2at% Cr. In the α-Fe(Zn) at position 6 the C and Mn decrease drastically to ~0.1 and 0.33 at% whereas the Si and Cr change only slightly. Typically, this additional austenite to ferrite transformation during cooling precedes in our system the position of the Zn diffusion front and almost everywhere the Zn containing α-Fe(Zn) and the bulk martensite are well separated by it, as can be seen in Fig. 9c.

The APT measurements showed that the Zn composition in the martensitic matrix is extremely low, in the order of 0.002 at%. In addition, at defects such as dislocations as well as at random grain boundaries no Zn segregation was found. It seems that not all, but only some specific PAGBs are significantly enriched and therefore weakened by Zn. Such PAGBs were identified in our AES und STEM measurements. A deep in-diffusion of Zn along an austenite GB (at high temperatures, transforms to martensite when quenched) from the α-Fe(Zn) into the bulk can only take place, if the Zn containing phase or liquid Zn is in direct contact with a suitable austenite GB.

At this point, we should discuss how a suitable austenite GB could look like. Typically, the solubility of Zn in austenite is quite low and Zn immediately forms α-Fe(Zn). Thus, one could expect, that deep in-diffusion is unlikely to occur due to this reaction with the surrounding matrix. However, another case could be, if Zn-rich austenite is stable and the α-Fe(Zn) formation does not take place. In this case still a limited solubility (strongly depend on the exact alloy element concentrations) of Zn in the austenite and a small bulk diffusion will be present, leading to a deeper diffusion of Zn along the GB compared to the case with α-Fe(Zn) formation. C and Mn are both well-known austenite-stabilizing elements. And indeed we find Mn enrichments at the position of the wetted PAGB, (Fig. 9d) which we assume to be the reason for a suppressed α-Fe(Zn) formation. This implies additionally that the austenite to Zn-free-ferrite formation cannot be present in these specific regions. The PAGB shown in Fig. 9b and 10 is an example of this behavior, where Zn penetrates several µm into the base steel. At this specific PAGB, the Zn-free ferrite layer is replaced by a region of Zn-rich martensite surrounding the PAGB.

The Zn-rich martensite (originated from Zn-rich austenite) with ~6at% Zn directly touch the α-Fe(Zn) phase. We see at this position a well visible indication of an increased Mn content in the surrounding of the Zn-modified PAGB (Fig. 10d). Unfortunately, no statement can be concluded for the C-distribution with STEM-EDX measurements.

Our results yield a strong hint that the exact local composition (e.g. enrichments or depletions of the alloy elements C, Mn and Si) at the interface region and the austenite GBs strongly influences the Zn-Fe reaction and the ferrite and martensite formation. This can lead to arrangement in which Zn-containing phases are in contact to wettable austenite GBs above the austenitizing temperature, where the Zn-Fe reaction is delayed or suppressed. Generally, we would expect that a Zn-enriched GB at austenitization temperature later leads to a Zn-enrich PAGBs. The AES measurements at fractured surfaces of the electrogalvanized, undeformed sample prove that the Zn penetration depth along PAGBs can reach 25 µm. Our experiments show that the actual Zn concentrations are very low, in the order of a monolayer or below, so that it is not possible to speak of bulk phases like α-Fe(Zn) or α′-Fe(Zn), which leads in some LMIE models to the cracking of the GBs. These PAGBs are weakened by these low amounts of Zn. From our experiments, we can estimate a number of wetted GBs in our system. From the number of cracks in Fig. 1c we can deduce a max. distance of 35µm between two cracks, which lead with prior austenite grain sizes of ~10µm to one crack at least for every 4$^{th}$ PAGB. From the presented TEM lamella measurements in Fig. 9 and additional measurements we can conclude that maximal every second GB is wetted. Of course, this is only a rough estimate, but leads to the insight that in our experiment Zn affects only 25%-50% of the PAGBs. During forming in the tool, these weak points can rip open and be the starting points of the macroscopic MIE cracks. After the first cracking larger amounts of liquid Zn or Zn vapor, can fill the initial crack and induce further cracking.

From the results, we can derive a simple model of the processes occurring during the initial stage of Zn metal embrittlement in a PHS steel:

1) During annealing above 870°C, the following phases are formed from bulk to surface: γ-Fe, α-Fe(Zn), liquid Zn/Fe, solid Zn-oxides (see Fig. 2 and 6).
2) The Zn and Fe interdiffusion triggers the growth of α-Fe(Zn), which separates the liquid Zn from the steel base. The growth is mainly determined by the α-Fe(Zn) Zn-Fe reaction respectively bulk diffusion. The GB diffusion is relatively quickly smeared out by the high solubility of Zn in the α-Fe(Zn).
3) Diverging diffusion gradients and different solubilities of the alloy elements in α-Fe(Zn) and γ-Fe lead to an additional region between the α-Fe(Zn) and the base steel. For our experiments this region transforms later during quenching into Zn-free ferrite, forming an almost continuous layer.
4) Only at positions where the α-Fe(Zn) Zn-Fe reaction is hampered by stabilized austenite, fast diffusion along austenite GBs can happen. We assume a generally much slower bulk diffusion of Zn in austenite due to lower solubility compared to ferrite. This leads to several µm deep penetration of Zn with a low concentration along these specific GBs (see Fig. 8-10) during annealing above austenitizing temperature.
5) During forming, when a force is applied, cracks are formed in the Zn coating at the position of the Zn weakened PAGB and expose the Zn sensitive GBs.
6) Present liquid Zn or Zn vapor covers the new surfaces and can further trigger MIE and macroscopic cracks are formed along these PAGBs (see Fig. 3).

7) This process can continue during hot forming in the presence of Zn until the hardening and the austenite-martensite transformation takes place.
8) Still present liquid Zn or Zn vapor accumulates in the microcrack tip via capillary effect (see Fig. 4 and 5) and can form additional Fe-Zn phases.

# 5. Conclusion

In our study, we investigated MIE and its underlying mechanisms by a detailed microstructure characterization on two different samples of the same 20MnB8 steels. Cracks in hot-dip galvanized, hot formed (starting temperature 600°C) and press hardened samples were investigated by AES, SEM-EDX and STEM-EDX. Zn containing phases in wedge shape smaller than 100nm were found at the crack tips at which Zn penetrated further into the base material, showing the importance of a present Zn source for MIE crack propagation. Additionally, it was showed via EBSD that the microcracks are formed at prior austenite grain boundaries.

On the electrogalvanized sample featuring the same thermal history without deformation we could detect Zn wetting of prior austenite grain boundaries with AES, TEM and STEM-EDX. Under ultra-high vacuum condition of a AES, Zn could be detected down to a depth of 25µm below the α-Fe(Zn) phases on *in situ* cracked martensite surfaces. The amount of Zn in the grain boundaries was estimated to be in the order of a monolayer Zn or less. We could additionally show that a low amount of Zn can wet austenite GBs during austenitizing annealing several tens of µm deep into the base steel matrix. Via APT we could determine an extremely low Zn composition of only 0.002 at% of the martensitic matrix, proofing that the detected Zn signal in STEM-EDX and AES at fracture surfaces must indeed stem from Zn wetted GBs and not the matrix.

We believe that austenite GB wetting happens at position, where the α-Fe(Zn)-ferrite reaction is suppressed and γ-Fe(Zn) can be formed. Due to a reduced diffusion of Zn in γ-Fe(Zn), extended diffusion along austenite GBs is then possible. Thus, if high concentration Zn-phases or liquid Zn comes into direct contact to suitable austenite GBs efficient diffusion along the GBs is possible if the formation of Zn-ferrite is hampered. From the above-described mechanism, it is clear that not all austenite GBs are wetted with Zn. We think that the C distribution are one major reason (but not measurable with STEM-EDX) as well as the Mn distribution for the suppression of the α-Fe(Zn) formation. This leads to Zn weakened GBs during austenitization and/or hot forming which are the origin of MIE cracking during forming and under stress.

Our study shows the importance to investigate the first stages of embrittlement, including the diffusion of alloy elements and austenite/ferrite/martensite transformation behavior. Experimental investigation of this complex interplay of different phenomenon is additionally complicated by experimental challenges to identify suitable regions for TEM and APT. These regions are difficult to identify due to their low Zn-concentrations, especially if it is currently not even clear, which GBs are wetted by Zn. Therefore, we are planning to perform correlative EBSD/APT/TEM measurements where TEM data could be acquired directly on APT tips prior to APT measurements. In further studies we want to increase the annealing times and investigate the diffusion along the grain boundaries, including also HR-STEM imaging. Especially the exact structure of the GBs seems to have a strong impact on the phase transformations near the GB and the diffusion along the GB and thus its impact for embrittlement. This includes the questions, how the Zn enrichment at the austenite GB can be described at different temperatures: as α-Fe(Zn), Γ-Fe(Zn) or Γ'-Fe(Zn), GB complexion or only as single Zn atoms forming a wetting and already leading to an substantial weakening of the GB.

# 6. Acknowledgments

The financial support by the Austrian Federal Ministry for Digital and Economic Affairs, the National Foundation for Research, Technology, and Development, and the Christian Doppler Research Association is gratefully acknowledged. Financial support from the Austrian Government and the provinces of Upper and Lower Austria within the COMET Program managed by the Austrian Research Promotion Agency (FFG) is gratefully acknowledged. The authors are grateful to U. Tezins and A. Sturm for their support to the FIB and APT facilities at MPIE. Baptiste Gault is acknowledged for helpful discussions about APT data analysis.

# Author contributions

**M. Arndt:** Conceptualization, Formal analysis, Investigation, Writing - Original Draft, Visualization **T. Truglas:** Investigation, Writing - Review & Editing **P. Kürnsteiner:** Formal analysis, Investigation, Writing - Review & Editing, Visualization **J. Duchoslavc:** Resources **K. Hingerl:** Conceptualization, Writing - Review & Editing **D. Stifter:** Resources, Writing - Review & Editing **C. Commenda:** Formal analysis, Investigation **J. Haslmayr:** Investigation S. Kolnberger: Conceptualization **J. Faderl:** Conceptualization, Resources, Writing - Review & Editing, Project administration **H. Groiss:** Conceptualization, Validation, Investigation, Resources, Data Curation, Writing - Original Draft, Visualization, Supervision, Project administration, Funding acquisition

# Supplementary Information to:

# Nanoscale Investigation of Microcracks and Grain Boundary Wetting in Press Hardened Galvanized 20MnB8 Steel

M. Arndt[a], P. Kürnsteiner[b,e], T. Truglas[b], J. Duchoslav[c, d], K. Hingerl[c], D. Stifter[c], C. Commenda[a], J. Haslmayr[a], S. Kolnberger[a], J. Faderl[a], H. Groiss[b*]

[a]voestalpine Stahl GmbH, voestalpine-Straße 3, 4020 Linz, Austria

[b]Christian Doppler Laboratory for Nanoscale Phase Transformations, Center for Surface and Nanoanalytics (ZONA), Johannes Kepler University Linz, Altenberger Straße 69, 4040 Linz, Austria

[c]Center for Surface and Nanoanalytics (ZONA), Johannes Kepler University Linz, Altenberger Straße 69, 4040 Linz, Austria

[d]Centre for Electrochemistry and Surface Technology GmbH (CEST), Viktor Kaplan Straße 2, 2700 Wiener Neustadt, Österreich

[e]Department Microstructure Physics and Alloy Design, Max-Planck-Institut für Eisenforschung GmbH, Max-Planck-Straße 1, 40237 Düsseldorf, Germany

*: corresponding author: heiko.groiss@jku.at

Supplementary figure S1:

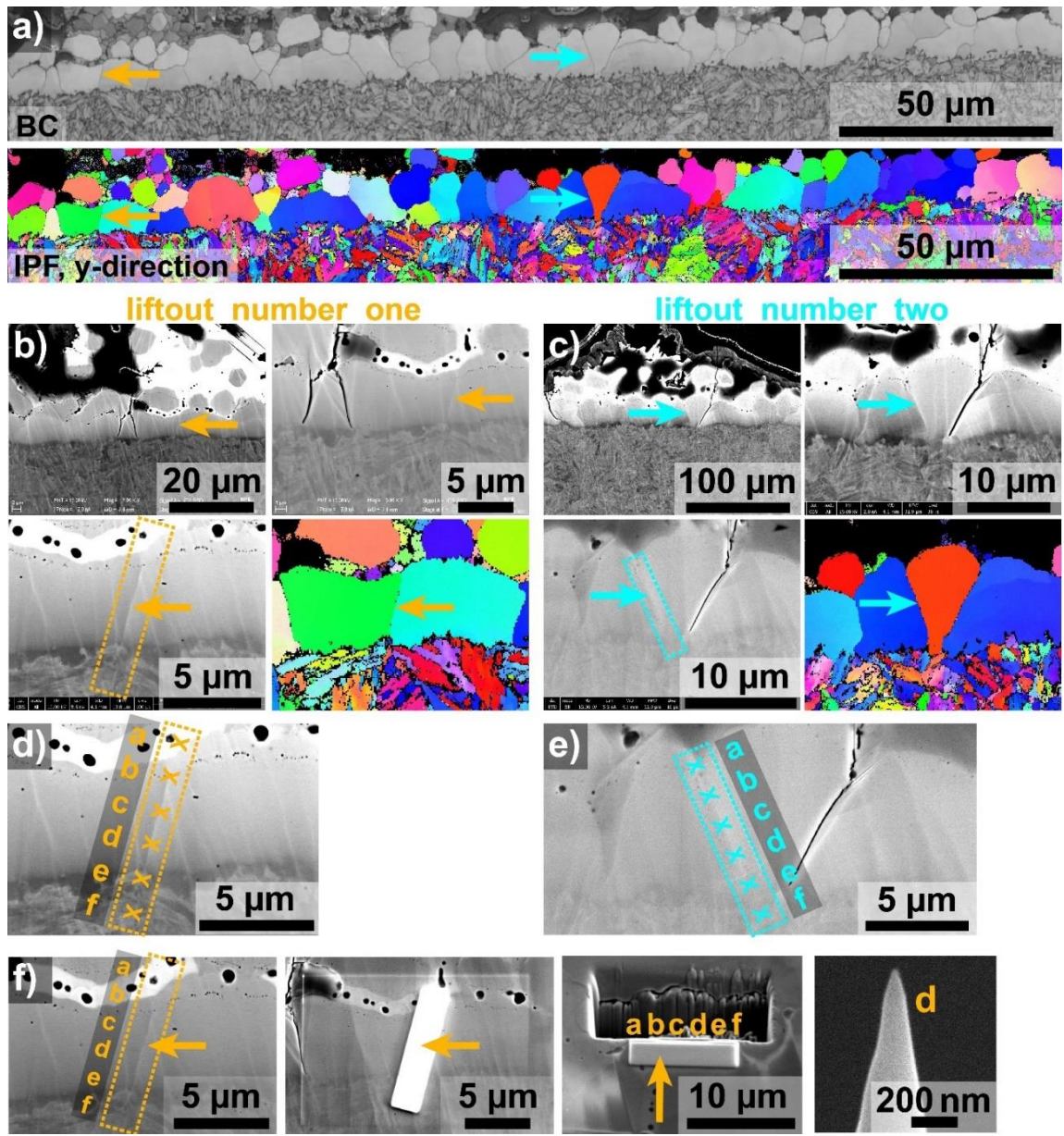

*Figure S1. This figure illustrates the atom probe tomography (APT) liftout procedure and how composition profile graphs were obtained from multiple APT measurements. a) shows a large-scale EBSD mapping of the interface region between Zn coating (top) and martensitic base steel (bottom). A band contrast (BC) image (top), as well as an inverse pole figure (IPF) map (bottom) are shown in a). The projection axis for the IPF map is perpendicular to the steel/coating interface (i.e. y-direction). Such mappings were used to screen a large number of potential grain boundaries and identify suitable ones for an APT liftout. We targeted grain boundaries that lead straight into the martensite so that we would be able to follow the GB into the martensitic microstructure and contain a GB within the APT tip (i.e. we targeted the arrangement shown in Fig. 8 b where the Zn containing α-Fe(Zn) directly touches the martensite without the occurrence of the Zn-free ferrite). Two GBs along which we performed a liftout are marked by orange and cyan arrows. b) and d) show the grain boundary that was lifted out for liftout number one (orange highlighting color) at different magnifications in backscattered electron (BSE) micrographs as well as the corresponding zoom to the IPF EBSD mapping, c) and e) show the grain boundary of liftout number two (cyan highlighting color). d) and e) mark the positions of the individual APT tips that were produced from the liftout. Each cross represents an individual APT tip, hence an individual APT measurement that yields a chemical composition at the position of the cross. The results of the chemical composition analysis are shown in the form of a graph in Fig. S3 and S4. and in tabular form in table S1 and S2. f) shows a few snapshots of the liftout procedure performed: first a promising GB is identified. A rectangular region around the GB is covered with a protective Pt deposit to preserve the GB during the subsequent FIB milling. Material is removed adjacent and underneath the Pt deposit using FIB milling. The remaining wedge-shaped material is lifted out using a micromanipulator, chopped into individual segments (here 6 segments) which are placed on Si posts and sharpened into needle shaped APT tips by annular FIB milling.*

Supplementary figure S2:

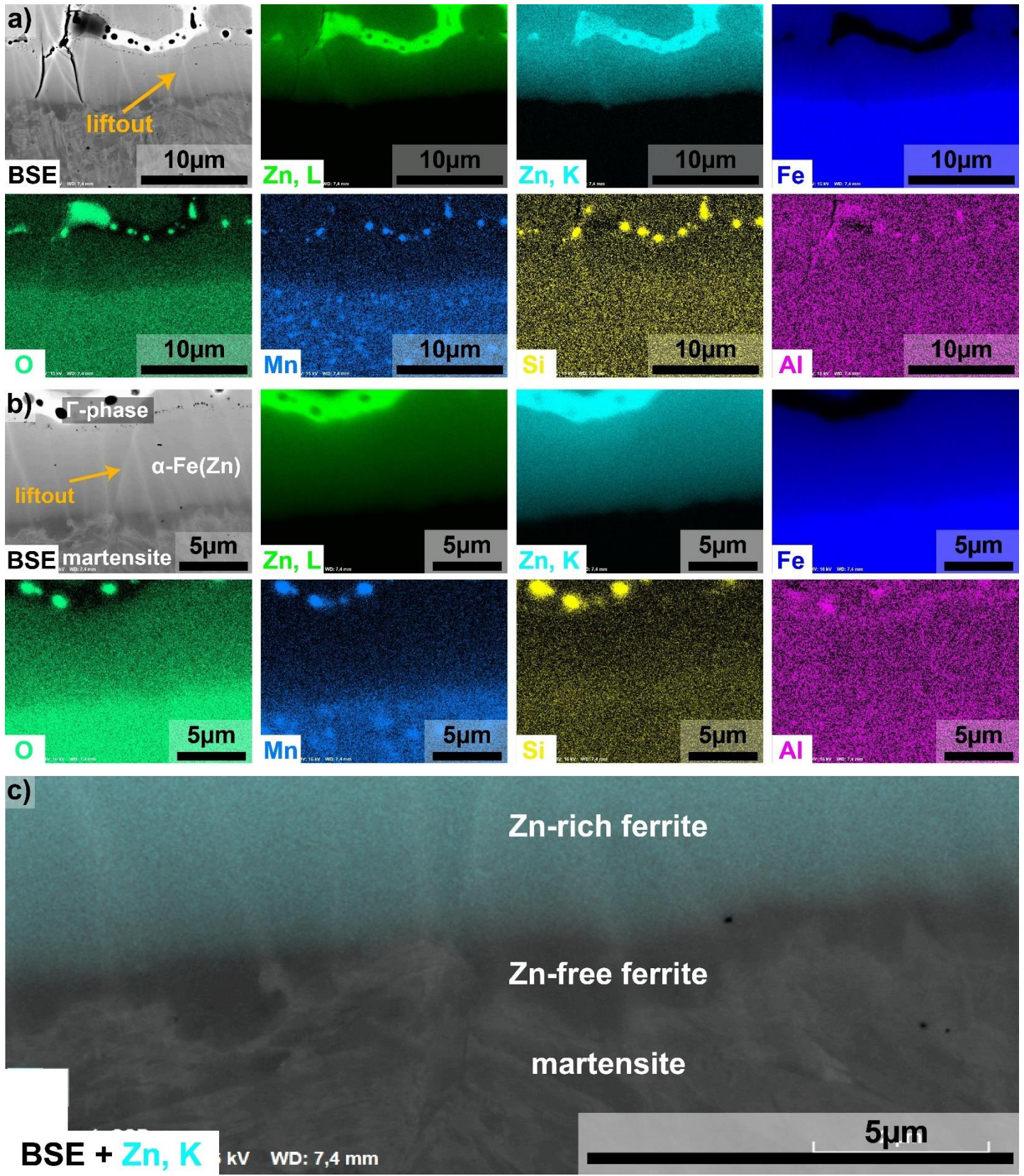

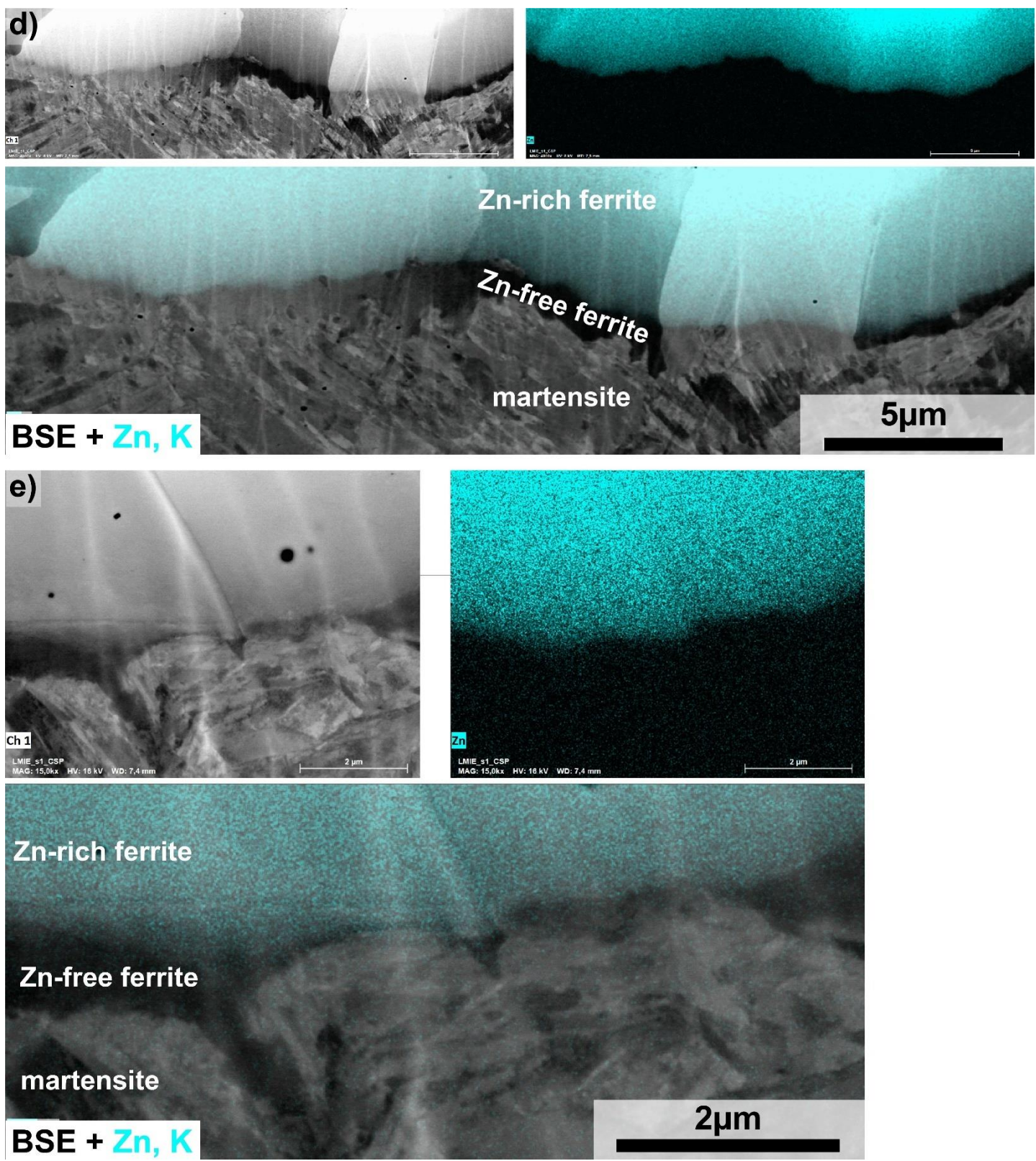

*Figure S2. EDX mapping of the boundary lifted out for the APT measurement set number one at different magnifications. a) covers a larger area of the Zn coating region. b) is a magnified view of the liftout region. Mappings using both L and K line are shown for Zn. K lines are a used for all other elements. The electron micrograph was acquired using backscattered electrons (BSE). Different features are discernable in the elemental mappings: α-Fe(Zn) at the top, a band of Γ-phase in the region of α-Fe(Zn), the martensitic base steel at the bottom and Si-Mn-oxides inside the Γ-phase. These Si-Mn-oxide particles seem also slightly enriched in Al. There appear to be some Mn-rich particles in the martensitic base steel, but it seems that the resolution of the SEM-EDX mapping here is not sufficient to clearly resolve them. c) shows an overlay of the BSE micrograph with the Zn EDX signal clearly demonstrating the layer of Zn-free ferrite separating the martensitic base steel from the Zn-rich ferrite (α-Fe(Zn) phase). d) and e) depict two additional examples Zn EDS mappings overlaid with BSE micrographs.*

Supplementary figure S3:

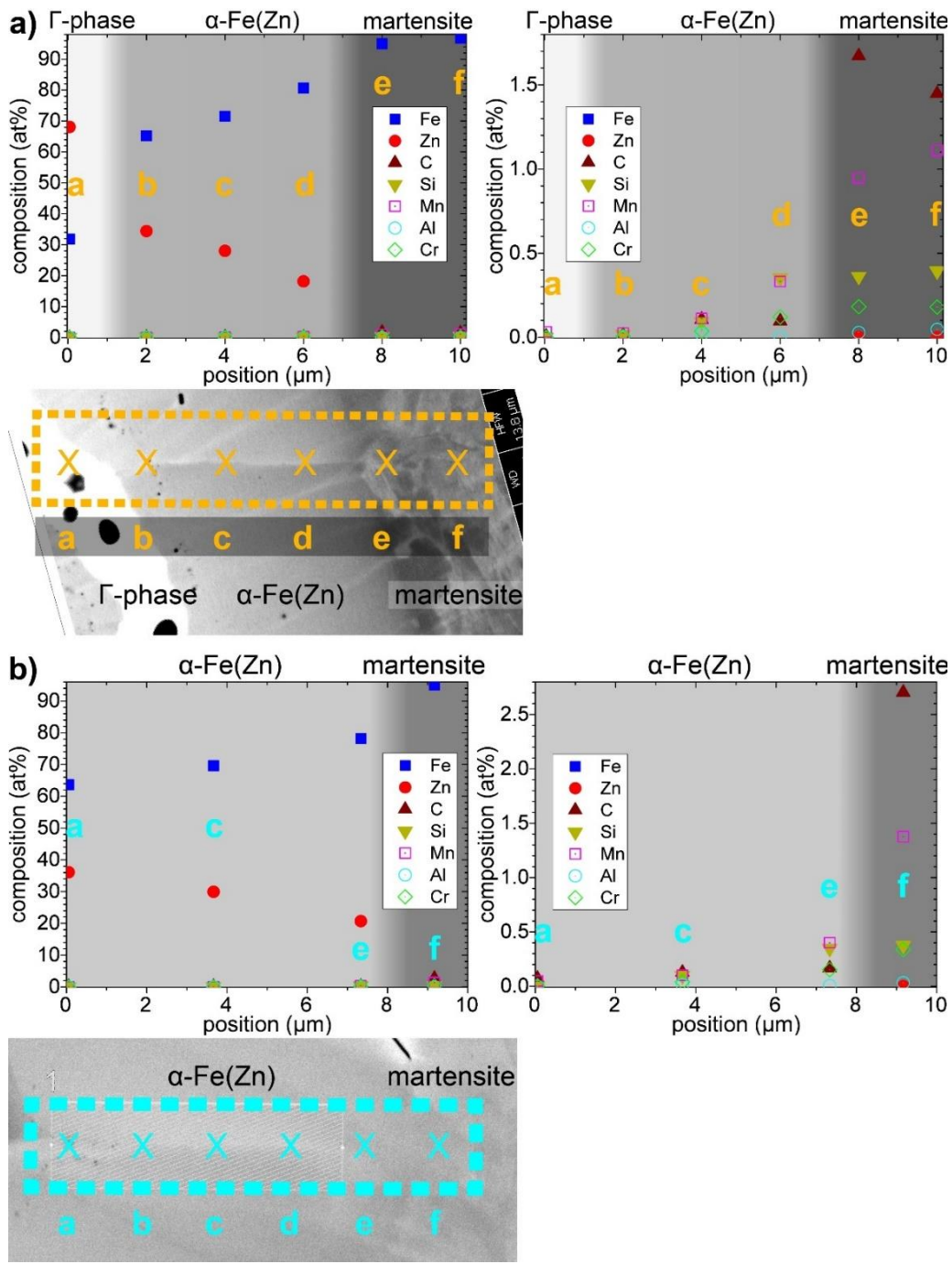

*Figure S3. Composition profiles along the Zn coating into the martensitic base steel (from left to right). a) shows the chemical composition of the 6 APT measurements from liftout number one (see Fig. S1 and S2). The electron micrograph underneath the graph is aligned in such way that the markings of the APT tip positions on the micrograph are aligned with the data points in the graph. Already from the micrograph it is apparent that the first APT measurement (number 1) is placed inside the Γ-phase, the next three APT measurements (number 2-4) are located inside α-Fe(Zn) while the last two measurement (number 5, 6) are in the martensitic base steel. b) is the same as a) but for liftout number two. Unfortunately, two APT tips (number 2 and 4) fractured prematurely, which is why the corresponding data points in the graph are missing. Here only APT tip number 6 is located in the martensitic base steel, all other APT tips are within the α-Fe(Zn) phase. The results of liftout number two are shown here, to demonstrate the reproducibility of the results obtained despite the small volume analyzed in each individual APT measurement. The same gradient of decreasing Zn concentration and increasing Mn, Si and Cr gradient moving towards the base steel is found in both sets of measurements. The C composition of the martensite is higher than the nominal composition of approximately 1 at% in both measurements in liftout one as well as the single measurement in this region in liftout two, which could be an enrichment of C during the α-Fe(Zn) formation on the steel base side due to a low solubility of the formed phase. Other reasons could be due to local variations in C concentration, as well as C contaminations during any of the preparation steps.*

## Supplementary table S1

*Table S1. Composition of the individual APT measurements from liftouts number one corresponding to the graph shown in Fig. S3 a).*

| liftout number one | | | | | | | | |
|---|---|---|---|---|---|---|---|---|
| measurement number | distance (µm) | Fe (at%) | Zn (at%) | Si (at%) | Mn (at%) | C (at%) | Al (at%) | Cr (at%) |
| 1 | 0 | 31.862 | 68.069 | 0.000 | 0.033 | 0.007 | 0.000 | 0.000 |
| 2 | 2 | 65.250 | 34.438 | 0.016 | 0.025 | 0.012 | 0.000 | 0.005 |
| 3 | 4 | 71.535 | 28.027 | 0.092 | 0.112 | 0.106 | 0.019 | 0.037 |
| 4 | 6 | 80.713 | 18.225 | 0.353 | 0.332 | 0.097 | 0.001 | 0.121 |
| 5 | 8 | 95.00 | 0.003 | 0.362 | 0.948 | 1.672 | 0.029 | 0.182 |
| 6 | 10 | 96.761 | 0.002 | 0.393 | 1.110 | 1.448 | 0.048 | 0.180 |

## Supplementary table S2

*Table S2. Composition of the individual APT measurements from liftout number two corresponding to the graph shown in Fig. S3 b). Note that the measurements number 2 and 4 from this liftout did not yield a result since the APT tips fractured prematurely.*

| liftout number two | | | | | | | | |
|---|---|---|---|---|---|---|---|---|
| measurement number | distance (µm) | Fe (at%) | Zn (at%) | Si (at%) | Mn (at%) | C (at%) | Al (at%) | Cr (at%) |
| 1 | 0.1 | 63.685 | 36.136 | 0.001 | 0.049 | 0.072 | 0.000 | 0.003 |
| 2 | 1.8 | -- | -- | -- | -- | -- | -- | -- |
| 3 | 3.7 | 69.639 | 29.937 | 0.097 | 0.102 | 0.127 | 0.013 | 0.039 |
| 4 | 5.5 | -- | -- | -- | -- | -- | -- | -- |
| 5 | 7.3 | 78.165 | 20.726 | 0.348 | 0.399 | 0.166 | 0.007 | 0.151 |
| 6 | 9.2 | 95.042 | 0.002 | 0.374 | 1.376 | 2.704 | 0.033 | 0.337 |

Supplementary figure S4:

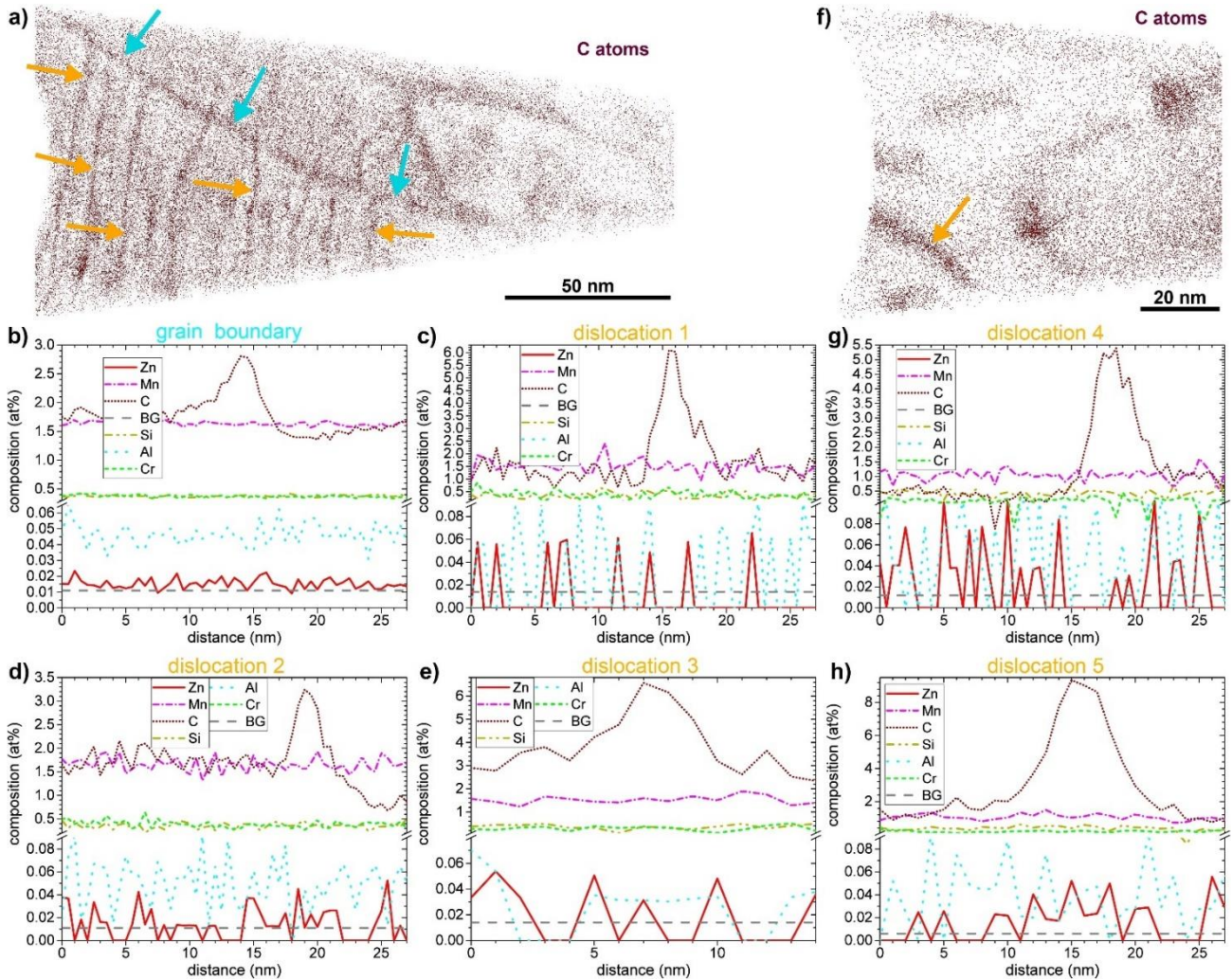

*Figure S4. APT analysis of segregation to grain boundaries and dislocations. The top of this figure shows so-called atom maps (i.e. all C atoms in a thin slice through the dataset) of two different measurements in the martensitic region. A GB is marked by cyan arrows, some dislocations are marked by orange arrows. b) corresponds to the composition profile through the boundary marked by cyan arrows in a). The composition profile was calculated along a cylinder of 20 nm diameter across a straight segment of the boundary. Note that the Zn composition in b) is higher than the aforementioned 0.002 at% which is due to the fact, that the Zn composition given for the martensitic matrix was corrected for the background, but for the graph we decided to show the uncorrected values together with the background (BG) in the graph. The BG was determined separately for each composition profile. c)-e) show profiles through different dislocation from the APT dataset in a). g) and h) show profiles through two different dislocations from the APT dataset in d). The atom map shown in a) corresponds to the reconstruction of APT tip number 6 from liftout number 1. The graphs in b), c), f) and g) are extracted from this dataset. The atom map shown in d) corresponds to the reconstruction of APT tip number 6 from liftout number 2. The graphs shown in e) and h) are extracted from this dataset.*

*It is apparent that the dislocations as well as the GB are marked by strong C enrichments. However, no other elements show any segregation tendency to the GB or dislocations. For the APT liftouts, we target grain boundaries in the α-Fe(Zn) that continued straight into the martensite and where no Zn-free ferrite section was present (see Fig. S1 and S2). Unfortunately, we could not unambiguously identify a PAGB in the APT measurements. The GB marked by cyan arrows in a) in the martensitic region is visible due to its C enrichment. However, no Zn segregation is present at this GB (as apparent in b)). Due to missing crystallographic information, we cannot determine whether this GB is a PAGB which does not show any Zn wetting or if it is simply a random martensite boundary. We think that usually not all PAGBs are covered with Zn and due to the difficult pre-identification of potential Zn-wetted GBs, there is only a low probability of finding a Zn-wetted PAGB within APT tips with a limited number of attempts.*